# A LARGE SCALE STRUCTURE AT REDSHIFT 1.71 IN THE LOCKMAN HOLE

J. Patrick Henry[1], Kentaro Aoki[2], Alexis Finoguenov[3], Sotiria Fotopoulou[4], Günther Hasinger[1], Mara Salvato[5,6], Hyewon Suh[1] and Masayuki Tanaka[7]

[1]Institute for Astronomy, 2680 Woodlawn Drive, Honolulu, HI 96822, USA; henry@ifa.hawaii.edu
[2] Subaru Telescope, 650 North A'ohoku Place, Hilo, HI 96720, USA
[3]Department of Physics, University of Helsinki, Gustaf Hällströmin katu 2a, FI-00014 Helsinki, Finland
[4]Department of Astronomy, University of Geneva, Ch. d'Écogia 16, 1290 Versoix, Switzerland
[5]Max Planck Institut für Extraterrestrische Physik, Giessenbachstr., 85748 Garching, Germany
[6]Excellence Cluster, Boltzmannstr. 2, 85748 Garching, Germany
[7]National Astronomical Observatory of Japan, 2-21-1 Osawa, Mitaka, Tokyo 181-8588, Japan

## ABSTRACT

We previously identified LH146, a diffuse X-ray source in the Lockman Hole, as a galaxy cluster at redshift 1.753. The redshift was based on one spectroscopic value, buttressed by seven additional photometric redshifts. We here confirm the previous spectroscopic redshift and present concordant spectroscopic redshifts for an additional eight galaxies. The average of these nine redshifts is $1.714 \pm 0.012$ (error on mean). Scrutiny of the galaxy distribution in redshift and the plane of the sky shows that there are two concentrations of galaxies near the X-ray source. In addition there are three diffuse X-ray sources spread along the axis connecting the galaxy concentrations. LH146 is one of these three and lies approximately at the center of the two galaxy concentrations and the outer two diffuse X-ray sources. We thus conclude that LH146 is at the redshift initially reported but it is not a single virialized galaxy cluster as previously assumed. Rather it appears to mark the approximate center of a larger region containing more objects. For brevity we term all these objects and their alignments as large scale structure. The exact nature of LH146 itself remains unclear.

## 1. INTRODUCTION

Young galaxy clusters, those at $z \gtrsim 2$ whose populations are not yet dominated by massive galaxies no longer forming stars but rather contain many galaxies near the peak of their star formation, are of special interest. They provide direct information about the early evolution of galaxies and of the intracluster gas and thus how these cluster constituents obtained their present properties. In particular, observations at these redshifts may show how and when star formation in cluster galaxies was quenched. Similarly, such observations may show how the cluster gas was heated, when it was enriched by heavy elements and can help disentangle the competing effects of energy injection (heating) versus cooling, since the latter should not have had sufficient time to greatly modify the gas properties (Voit 2005).

Another reason to identify very distant galaxy clusters is the increased leverage provided by such objects compared to local ones in observational cosmology. Clusters provide regions containing baryon and dark matter fractions representative of the universe because very large volumes of material at the mean density of the universe are needed to encompass a cluster's mass. Thus observations of clusters permit the direct measurement of the evolution of these quantities. Allen, Evrard and Mantz (2011, Sections 2.2.2 and 4.2) discuss the application of this idea to cosmology. In addition, the evolution of cluster number density can be used to constrain cosmological parameters and the gaussianity of initial density fluctuations (Allen, Evrard and Mantz, 2011).

As sparse as our information is about the properties of young galaxy clusters, our observational information about their environment is even more lacking. Tadaki et al. (2012) present one such comprehensive studies around the $z = 1.62$ cluster ClG J0218.3-0510 (aka IRC0218 and XMM-LSS J02182-05102). This information is interesting because locally star formation has ceased in high density environments, but galaxies in the field are still actively forming stars. This dependance is still in place at $z \sim 1$ (Muzzin et al. 2012). By $z \sim 1.6$ however the relation may be weakening. For example Tadaki et al. find that star-forming [OII] emitters show an over density of >10 compared to the field in the densest core of the cluster they study. In addition, they find that the cluster is embedded in a large ~20 Mpc filamentary structure traced by [OII] emitting galaxies. Clearly much larger samples of young galaxy clusters are needed to clarify the situation.

To date there have been two ways of finding young galaxy clusters: by their extended X-ray emission and by clustering of IR galaxies. Reports of such objects at $z > 1.6$ with confirming spectroscopic redshifts include: Kurk et al. (2009), $z = 1.61$; Tanaka et al. (2013), $z = 1.61$; Tanaka, Finoguenov & Ueda (2010), Papovich et al. (2010), Pierre et al. (2012), $z = 1.62$; Muzzin et al. (2013), $z = 1.63$; Henry et al. (2010; H10 hereafter), $z = 1.75$; Stanford et al. (2012), $z = 1.75$; Zeimann et al. (2012), $z = 1.89$; Gobat et al. (2013), $z = 2.00$.

We report in this paper new observations in the field containing the object described in H10, which we named LH146. We increased the number of concordant spectroscopic redshifts from one to nine, find two galaxy overdensities approximately equidistant from LH146 (Section 2) with slightly different redshifts (Section 3) and find two more diffuse X-ray sources aligned on the same axis as the galaxy overdensities, one of which may be at the same redshift as LH146 (Section 4). Thus LH146 is not an isolated virialized cluster. Rather it appears to mark the approximate center of a larger region containing more objects. For brevity we term all these objects and their alignments as large scale structure. Kurk et al. (2009) report a similar large scale structure as ours at a similar redshift and we compare and contrast them in Section 6. Despite the new data, the exact nature of LH146 itself is unclear (Section 6).

The cosmology assumed in this paper has parameters $H_0 = 72$ km s$^{-1}$ Mpc$^{-1}$, $\Omega_m = 0.25$, and $\Omega_\Lambda = 0.75$. An object at redshift 1.75 is then at an angular diameter distance of 1736 Mpc implying an angular scale of 0.505 Mpc arcmin$^{-1}$; the look-back time to it is 9.9 Gyr or 3.9 Gyr since the Big Bang. We use AB magnitudes unless stated otherwise.

## 2. GALAXY DISTRIBUTION

In H10 we used a preliminary photometric redshift catalog for the Lockman Hole. The final catalog is now available (Fotopoulou et al. 2012) with substantially better redshift accuracy. Using 19 photometric bands from the far-ultraviolet to the mid-infrared, we achieve, for normal galaxies, a redshift accuracy of $\sigma_{\Delta z/(1+z)} = 0.036$ with 12.6% outliers. We identify galaxies in this paper by their ID numbers from the final catalog, so ID 171765 in H10, the only object with a spectroscopic redshift in that paper, is now ID 206708.

We plot in Figure 1 left galaxies near LH146 with photometric redshifts between 1.65 and 1.80 and with errors ≤ 0.1. The average photometric redshift 1 σ error for objects from these cuts is 0.046. Most of the galaxies appear to lie on a structure running SE to NW. Generating isopleths from the galaxy distribution reinforces this appearance and shows that there are two main concentrations at the SE and NW ends that we name LH146SEO and LH146NWO, respectively (see Figure 1 right). There also appears to be a linear galaxy overdensity between the two concentrations. We calculate the significance of LH146SEO, LH146NWO and the linear overdensity from the number of galaxies contained within the second lowest contours at the SE and NW and a non overlapping 1.2′ x 2.2′ rectangle between them. There are 13, 8 and 15 galaxies in regions 1.23, 0.65 and 2.65 arcmin$^2$, respectively. With a background of 2.08 arcmin$^{-2}$, the Poisson probabilities of obtaining the observed number of galaxies are 2.518 x 10$^{-6}$, 7.093 x 10$^{-5}$ and 3.984 x 10$^{-4}$, respectively. As a point of reference, the summed Poisson probabilities up to and including the observed counts equal the integral Gaussian probabilities up to and including 4.9σ, 4.2σ and 3.5σ, respectively. Note that we chose the photometric redshift bin size and center based on the spectroscopy described in Section 3. Thus only one trial enters the significance calculation since we did not peak up the clustering signal by adjusting those parameters. The linear overdensity could be either the superposition of galaxies from LH146SEO and LH146NWO, or from them and LH146, or a physical filament of galaxies, or any combination of these.

## 3. INFRARED DATA

### 3.1 Subaru Spectroscopy

We used two spectrographs mounted on the 8.3 m Subaru telescope on Mauna Kea. The first was MOIRCS (Ichikawa et al. 2006; Suzuki et al. 2008) in multi-object spectroscopy mode with the zJ500 grating and 0.8″ slits (resolution R ~ 500). The second was FMOS (Kimura et al. 2010) in cross-beam switching mode at H-short (1.4 - 1.6 μm) and H-long (1.6 - 1.8 μm) with high resolution (R ~ 2200). Cross-beam switching allocates two fibers separated by 90″ to each target, one of which is used for sky subtraction. Table 1 is the journal of the observations. We reduced the MOIRCS data with

MOIRCSMOSRED, a set of IDL scripts originally written by Yoiuchi Ohyama and revised by Tian-Tian Yuan. We describe our procedures in H10. We reduced the FMOS data using the publicly available IRAF-based software FIBRE-Pac (FMOS Image-Based REduction package; Iwamuro et al. 2012). FIBRE-Pac includes background subtraction, and corrections for detector cross talk, bias difference, bad pixels, spectral distortion, and the removal of residual airglow lines.

We confirmed our previous redshift for ID 206708. We placed six of the seven objects with concordant photometric redshifts from H10 on MOIRCS slits. Three were too faint to obtain a redshift while the photometric redshifts of two of the remaining three were confirmed. Their IDs are 88636 and 87566. ID 88011 is the object whose spectroscopic redshift (0.900) disagrees with its photometric redshift. In total we obtained eight new concordant spectroscopic redshifts. Table 2 summarizes the properties of these nine galaxies and we show their spectra in Figure 2. Not unexpectedly most objects for which we could obtain a redshift were emission line galaxies, but we did obtain two absorption line redshifts (ID 206708 and ID 87566).

There were three objects projected serendipitously on the sixteen slits implying a serendipitous fraction 3/16. We could obtain redshifts for two of them (ID 88683 and ID 94757). Note that we did not know the galaxy IDs nor their photozs when we designed the slit mask as the final catalog was still under construction. Thus the objects are truly serendipitous. Both of them had redshifts that placed them in the large scale structure but not because they were companions to targets selected to be members (targets were ID 88413 and ID 167248 with spectroscopic redshifts 1.112 and 1.381, respectively). The fraction of photometric redshifts between 0 and 2 in the field of Figure 1 in the slice 1.675 - 1.755 that places them in the large scale structure is 72/2215. Thus the probability that two serendipitous galaxies are in the large scale structure is 0.004092 from the binomial distribution for two successes in sixteen trials with a probability of a single success equal to 3/16 x 72/2215. As a point of reference, the summed Binomial probability for zero, one and two successes equals the integral Gaussian probability up to and including $3.7\sigma$, and is additional evidence that there is an over density of objects in that redshift slice in this field. Although a posteriori, this calculation is conservative since it assumes we can measure redshifts for all objects on the slits and that we can only measure redshifts less than 2.

We show histograms of photometric and spectroscopic redshifts in Figure 3. There is a peak in the photometric redshift distribution at $z \sim 1.7$ which the spectroscopic redshift distribution seems to resolve into two components. We indicate in Figure 1 left the positions of the galaxies in these two components. Although it is difficult to be certain with only nine galaxies, it appears that there is position-velocity structure. Objects to the SE of LH146 are at a lower redshift (colored blue in Figure 1 left) than those to the NW (colored red). The average redshift and velocity dispersion (calculated from the redshift dispersion with c $\sigma_z/(1+z)$) of the five SE galaxies is 1.685 and 966 km s$^{-1}$ while that of the four NW galaxies is 1.751 and 421 km s$^{-1}$. These quantities are 1.714 and 3921 km s$^{-1}$ for all nine galaxies. The gapper algorithm (Wilman et al., 2005) gives nearly the same results. The large velocity dispersion when taking all nine objects together is evidence that we are observing more than one object.

## 3.2 Spitzer Imaging

The Lockman Hole region was surveyed several times to varying depths by Spitzer. These include the Spitzer Wide-Area Extragalactic Survey (Lonsdale et al. 2003; SWIRE) with nominal exposure 4 x 30s, the Medium Survey with nominal exposure 5 x 100s, The Cosmic IR Background Survey with nominal exposures of 60 x 200s and 30 x 200s in different fields and the Spitzer Extragalactic Representative Volume Survey (Mauduit, et al, 2012; SERVS) with nominal exposure 12 x 100s. The first three of these were acquired during the cryogenic phase of the mission, they are coadded in The Spitzer Heritage Enhanced Image Products and we use only these data in what follows. Data at the region of interest to us comes from SWIRE and the Medium Survey for a nominal total of 620s. The individual fields are overlapped to provide contiguous coverage, so the nominal exposures are minimum values. Right at LH146 the coverage is 14.1 pointings according to the coverage file, but the separation into SWIRE and Medium Survey pointings is not known to us. Figure 4 is a three color image comprised of $B_J$, $z'$ and [3.6], which are our deepest bands on both sides of the 4000Å break relative to the expected magnitude of a $z = 1.71$ elliptical galaxy. Galaxies with redshifts placing them in the large scale structure are circled.

## 4. X-RAY AND RADIO SOURCE DISTRIBUTION

In H10 we concentrated on LH146. Motivated by the discussion in the previous two section, we zoom out from LH146 to ascertain whether the X-ray source distribution reflects in some way the galaxy distribution. Our X-ray data come from summing all three XMM instruments, which yielded an equivalent MOS1 exposure of 1.883 Ms at the location of interest. Figure 1 right shows the locations of XMM sources from Brunner et al. (2008), the exes in the figure. These sources are primarily point-like as Brunner et al. find only 13 of their 409 sources are extended. The green contours in Figure 1 right show the surface brightness distribution of X-ray diffuse sources found and characterized as described in H10. They come from an à trous wavelet reconstruction of a PSF subtracted residual image using the 32″ and 64″ scales where the points sources were detected on the 8″ and 16″ scales. This procedure is analogous to reconstructing an image using only some of its Fourier modes. Figure 4 shows a finder chart for these sources as ellipses approximating the lowest detectable X-ray surface brightness contour. We include all diffuse X-ray sources we have found in the fields of Figures 1 right and 4 to show they are unlikely to be source confused.

There are three diffuse X-ray sources with the same general morphology as the galaxy structures. These diffuse sources are aligned on nearly the same axis as LH146SEO and LH146NWO, two of them have nearly the same angular extent as LH146SEO and LH146NWO, and these two are separated by nearly the same amount as LH146SEO and LH146NWO. We denote the two at either ends as LH146SEX and LH146NWX (see Figure 1 right). The alignment and angular separation of these three sources is very similar to that of the independently determined galaxy overdensities, hence they may be part of the large scale structure formed by the galaxy distribution. However, until we identify the source of the X-rays from LH146NWX and LH146SEX we obviously cannot rule out that they are unrelated fore or background sources.

We note that the three green contours fully inside Figure 1 right appear to follow some of the Brunner et al. sources, possibly indicating systematic residual emission from those sources. Arguing against this interpretation are: 1. While the green contours follow the locations of the Brunner et al. sources, they do not follow their flux. Indeed the brightest Brunner et al. source in Figure 1 right, LH168 at $\alpha(2000) = 163.3825$ $\delta(2000) = 57.4151$, is > 5 times brighter than the second brightest such source and 10 times brighter than the third brightest but does not show residual emission. 2. The central diffuse source, LH146, is extended to the eye, i.e. is not completely a systematic effect.

There are two known classes of extragalactic diffuse X-ray sources. The first is the well - studied thermal radiation from galaxy groups and clusters. The second is non thermal radiation from radio sources whose electrons inverse Compton (IC) scatter seed photons up to X-ray energies. In the latter case the X-ray source will be significantly longer lived than the radio source since much lower energy electrons than those needed to produce detectable radio emission are sufficient to produce detectable X-ray emission. The cosmic microwave background is an attractive source of seed photons since the surface brightness produced by IC scattering of it is independent of redshift (Schwartz, 2002; termed ICCMB). Further such ICCMB sources are predicted to outnumber the thermal cluster sources for $z > 2$ and X-ray luminosities > $10^{44}$ erg s$^{-1}$ (Celotti & Fabian, 2004), although initial searches for these sources by one of us (AF) are finding a somewhat different redshift distribution from the predictions.

We examine whether the NW and SE diffuse X-ray sources belong to either of these two classes; first thermal sources. The X-ray emission from neither source can be thermal emission from gas trapped by the potential well that confines the galaxies since neither is spatially coincident with them. They can be virialized or partially virialized groups about to fall into those potential wells. The southern part of 146NWX could be a group. There are several similar brightness orange galaxies straddling the SW part of the ellipse indicating that source in Figure 4. The ID numbers of the brighter ones in the Fotopoulou et al. (2012) catalog are 91579, 92077, 92255, 92439, 92538, 92717, 92836 and 92856. These objects have red sequence colors at $z = 0.60$, their average photometric redshift.

Next we examine the IC non thermal source scenario. For this purpose we have also plotted in Figure 1 right the positions of radio sources from the very deep 1.4 GHz catalog of Biggs & Ivison (2006). There are only two places where a Biggs & Ivison radio source coincides with a Brunner et al. X-ray source, one each at the east and north edges respectively of LH146SEX and LH146NWX, making them prime candidates for the electrons needed to power ICCMB scattering. Croston et al. (2005) find that single lobe X-ray emitters occur in approximately 25% of double lobed radio sources, so our detections of only single X-ray lobes is not uncommon. The radio, X-ray and optical IDs from Biggs & Ivison, (2006), Brunner et al. (2008) and Fotopoulou et al. (2012) are 440, 197 and 95748 (NW) and 473, 116 and 206444 (SE), respectively. Their redshifts are, in the same NW - SE order, 1.51 - 1.68 (photometric 1$\sigma$ range) and 0.607 (spectroscopic). Thus in the IC interpretation, LH146NWX could be in the large scale structure given its photometric redshift range but LH146SEX is not.

Starting with the net count rate in the 0.5 - 2.0 keV energy band from the background and point source subtracted exposure divided image, we can determine fluxes, luminosities and other physical properties of LH146SEX and LH146NWX appropriate to the two classes. The counts are from simple aperture photometry of the corresponding green ellipses in Figure 4. For the thermal case we use scaling relations to infer various characteristic properties in an iterative process described in H10. These properties are given in Table 3. Fluxes and luminosities are corrected (extrapolated or interpolated) to $r_{500}$ using the beta model given in H10. The temperature and mass come from the luminosity - temperature and mass - luminosity relations used in H10. The equivalent velocity dispersion comes from the mass using $\sigma^2 = GM_{200}/3r_{200}$. Table 4 gives properties of possible ICCMB sources, assuming a typical photon index $\Gamma$ to obtain fluxes and luminosities. Although an ellipse centered on $\alpha(2000) = 163.3054$, $\delta(2000) = +57.4289$, with semimajor and semiminor axes = 19.5″ and 17.5″, and orientation 9.9º W of N would isolate the posited ICCMB source from the possible z = 0.60 group, it is premature to use that region to refine the fluxes and luminosities in Table 4 because we have not resolved LH146NWX into separate sources.

## 5. IS THERE A RED SEQUENCE IN THE LARGE SCALE STRUCTURE?

At z < 1.3 the red cluster galaxies lie on a very narrow sequence in a color-magnitude diagram with a color dispersion < 5%. This red sequence is a sequence of quiescent elliptical galaxy masses (Gallazzi et al., 2006), with the more massive (brighter) galaxies retaining more metals (redder). The slope and scatter about the mean red sequence of most clusters observed to date show little evolution because all stars in ellipticals formed long before the usually accessible observation epoch. This lack of evolution is expected to change at z ~ 2 because of the dramatic change clusters and the galaxies they contain are thought to undergo then. It is thought that the first clusters form around that epoch since the observed structures surrounding z > 2 radio galaxies are interpreted as being protoclusters (see the review by Miley & De Breuck, 2008). In addition many galaxies are assembled then and they are vigorously forming stars (e. g. Hopkins 2004; Daddi et al. 2007). Figure 10 of Gobat et al. (2011) is a nice graphical presentation of these ideas.

A quick inspection of Figure 4 shows that there does not appear to be the large overdensity of very red galaxies of the appropriate brightness as in the well-observed cluster MS1054-0321 at z = 0.831 (see Figure 2 of Blakeslee et al., 2006). Admittedly none of the other clusters near z = 1.75 are as spectacular as MS1054-0321 either, see Figure 2 of Tanaka, Finoguenov & Ueda (2010; ClG J0218.3-0510 z = 1.623), Figure 5 of Muzzin et al. (2013; SpARCS J022427-032354 z = 1.633), Figure 1 of Stanford et al. (2012; IDCS J1426.5+3508 z = 1.75), Figures 1 and 2 of Zeimann et al. (2012; IDCS J1433.2+3306 z = 1.89), and Figure 1 of Gobat et al. (2011; CL J1449+0856 z = 1.995) but all of them show more clumping of red galaxies than is apparent in Figure 4.

An observed red sequence can consist of both quiescent elliptical galaxies and dusty star-forming galaxies. Papovich et al. (2012) and Quadri et al. (2012) found the contamination of the red sequence from dust-obscured galaxies in the z = 1.623 cluster ClG J0218.3-0510 is ~20%. Any star-forming galaxies need to be removed to make comparisons with lower redshift clusters and to take advantage of the more easily interpreted ellipticals. Four of the five studies listed at the end of the last paragraph had morphological information from HST that could be used to make the distinction. We do not have HST observations and use our photometry. One might expect that the best fitting SED template from the photometric redshift determination would classify each object. This expectation is not realized because there is a degeneracy in the fits: extinguished starburst templates mimic unextinguished early type templates, exactly the degeneracy we are trying to break. Middelberg et al. (2013) find that, for a sample of morphologically selected early type or bulge-dominated galaxies, the best fit SED is an extinguished starburst 50% of the time. Instead we use a $(U-V)_{z=0}$ - $(V-J)_{z=0}$ color-color diagram to isolate quiescent galaxies in the manner of Wuyts et al. (2007) and Williams et al. (2009), as shown in Figure 5. Here $U_{z=0}$, $V_{z=0}$ and $J_{z=0}$ are the rest frame magnitudes interpolated from the observed $z'$, J and [3.6] magnitudes using the InterRest scripts (Taylor et al., 2009). The redshifted U, V and J effective wavelengths are quite close to the observed effective wavelengths of $z'$, J and [3.6] and we use these latter three bands for the interpolation. Filter curves are from Bessel (1990; U, B, V), Tokunaga, Simons & Vacca (2002; $J_{z=0}$), Hewett et al. (2006, J observed) and Capak (2013, $z'$ and [3.6]). The line in Figure 5 shows the Williams et al. (2009) cuts above which are the quiescent ellipticals. Using the observed $z'$, J and [3.6] and the cuts of equation (2) of Papovich et al. (2012) selects the same set of quiescent galaxies. The ID numbers of the quiescent galaxies in the Fotopoulou et al. (2012) catalog are 90119, 90495, 187968, 187994 and 206708.

In Figure 6 we show our $(U-B)_{z=0}$ - $M_{B,z=0}$ (Vega) color-magnitude diagram where $B_{z=0}$ is the rest frame B magnitude interpolated from the observed $z'$, J and [3.6] as above. We over plot the red sequence of ClG J0218.3-0510 at z = 1.62 (Papovich et al., 2010; Figures 6 and 7), which is ~ 0.15 magnitudes bluer than most of our quiescent galaxies. The colors of the red sequences in four cluster at z ~ 1.2 in Table 2 of Mei et al. (2009) are also separated by 0.15 magnitudes, perhaps indicating a similar spread in zero points (or age) at z ~ 1.2 and ~ 1.7. The dispersion in the $(U-B)_{z=0}$ color for the five quiescent galaxies is 0.13, which is equal to the average color error making it an upper limit to the intrinsic color dispersion.

Figure 6 also shows theoretical red sequences of galaxies that formed at $z_f$ = 2, 3, and 10. We use the simple stellar population (SSP) evolution models from Bruzual & Charlot (2003) with the Chabrier initial mass function (Chabrier 2003) and no dust extinction. We determine the zero points of the three models by varying their metallicities to fit the red sequence of the Coma cluster (Bower et al. 1992). All but one of the quiescent galaxies have a $z_f \approx 10$ or an age of 3.5 Gyr, although these values are really lower limits since the colors in Figure 6 do not change much for higher formation redshifts. This age is similar to red sequence galaxies in MS1054-0321 at z = 0.831 where Blakeslee et al. (2006) find $z_f$ = 1.6 to 3.2 or ages 2 to 5 Gyr, in IDCS J1426.5+3508 at z = 1.75 where Stanford et al. (2012) find $z_f$ = 5 to 6 or ages of 2.7 to 2.9 Gyr and in IDCS J1433.2+3306 at z = 1.89 where Zeimann et al. (2012) find $z_f$ = 3.5 to 6 or ages of 1.7 to 2.6 Gyr.

Thus LH146 has 5 red sequence galaxies with ages similar to other such high redshift cluster galaxies. As a sanity check, how many such galaxies are in other clusters at similar redshifts? The highest fidelity comparison we can make is with the data in Figure 6 of Papovich et al. (2010) for ClG J0218.3-0510 at z = 1.62. Their quiescent galaxies are selected with color cuts that yield our 5 galaxies, with photometric redshift cuts similar to ours and we have deep UKIDSS data for both samples. There are 7 ClG J0218.3 objects brighter than the faintest LH146 quiescent galaxy (J < 21.9 Vega) with a probability > 0.75 that their photoz is at the cluster redshift. This photoz probability cut yields galaxies with a photoz accuracy of ~0.07 (from Figure 5 of Papovich et al., 2010), close to our average photoz accuracy of 0.05 after our photoz accuracy cut. Comparison to other clusters with redshifts similar to ours is difficult because the published data mostly come from HST with different filters and with very different spatial resolution. A differential comparison that ameliorates these difficulties is to compare the number of red sequence galaxies no fainter than 1.5 magnitudes from the brightest quiescent galaxy, the flux limit of our data. In this magnitude interval there are 5 red sequence galaxies in MS1054-0321 (Figure 1 of Mei et al., 2009), 4 in IDCS J1426.5+3508 (Figure 6 of Stanford et al., 2012), 8 in IDCS J1433.2+3306 (Table 2 of Zeimann et al., 2012) and 12 in ClG J0218.3-0510 (Figure 2 of Rudnick et al., 2012). In so far as we can compare the clusters, there are similar numbers of red sequence galaxies.

The boxed galaxies in Figure 1 left mark the locations of the quiescent galaxies. They are not found predominately at or near the centers of the galaxy overdensities as at low redshifts. This distribution is reminiscent of red sequence galaxies in the $z = 1.89$ cluster IDCS J1433.2+3306, which are on two parallel chains (see Figure 2 left of Zeimann et al., 2012) and those in the $z = 0.831$ cluster MS1054-0321, which are on a single chain (see Figure 2 of Blakeslee et al., 2006).

So the answer to the question posed by the title of this section is yes since we can identify a similar number of galaxies with similar colors, ages and spatial distributions to those identified as red sequence members in some other high redshift clusters. Since we only have photozs for four of the five objects, we cannot completely rule out the possibility that they are superpositions but the spectroscopy presented here make that unlikely (compare columns 4 and 5 of Table 2).

## 6. DISCUSSION

We based our original identification of LH146 as a redshift 1.753 galaxy cluster on only one spectroscopic redshift. The most significant result reported here is obtaining another eight spectroscopic redshifts that confirm and extend our previous result. All redshifts do not seem to belong to a single Gaussian distribution, but appear to be distributed about at least two means. There are now four redshifts associated with the original value averaging 1.751, and the other five averaging 1.685. Further, there is a segregation on the plane of the sky with galaxies with lower redshifts to the SE and those with higher values to the NW. The distribution of galaxies on the plane of the sky in a photometric redshift slice encompassing the two spectroscopic averages shows a similar distribution with two significant concentrations oriented SE - NW. A line connecting the SE and NW galaxy peaks passes within 12″ or 100 kpc to the South of the LH146 peak surface brightness. Similarly, a line connecting the SE and NW X-ray surface brightness peaks is parallel to the former line and also passes within 12″ or 100 kpc of the LH146 peak, but to the North. The similar orientations of the three galaxy concentrations and the three diffuse X-ray sources argues that they are associated.

However these outer extended X-ray sources have not been conclusively identified, so that association must be tentative. The strongest candidate association is LH146NWX that we identify as a ghost radio lobe inverse Compton scattering the CMB. The central radio/X-ray nucleus has a concordant photometric redshift. Both the nuclear radio flux and extended X-ray flux are ~5x scaled down versions HDF130, another inverse Compton ghost source at the similar $z = 1.99$, and their linear sizes are similar (Fabian et al., 2009). This object may be an example of preheating of gas before it falls into a cluster potential well (e. g. Ma et al., 2013). Such preheating is postulated to break the expected self similarity of galaxy cluster X-ray properties.

It is difficult to characterize precisely the totality of what is in this field because we do not know the exact nature of the optical and X-ray components nor of the relations, if any, among them. We do not know whether either of the galaxy concentrations have collapsed, i.e. whether they are more properly classified as protoclusters. There may be only one mass concentration with the others destined to merge with it, but if so we do not know which it is because the velocity difference between the two spectroscopic redshift averages is ~7300 km s$^{-1}$, much larger than can be induced by any mass we have identified. The separation of the galaxy peaks on the sky is 3.2′ or 1.6 Mpc. Their separation along the line of sight is 42 Mpc at z = 1.75 (82 Mpc comoving) if their redshift difference is pure Hubble flow. Such separations are commonly seen in numerical simulations of large scale structure (e.g. Springel et al., 2005). Kurk et al. (2009) find a similar structure at z = 1.61, compare their Figures 2 and 1 with our 1 left and 3. They find two galaxy concentrations separated by 4.5′ connected by a filament of galaxies and two redshift peaks, but only separated by 0.008. The Kurk et al. structure thus appears to be perpendicular to the line of sight, while we seem to be viewing the LH146 large scale structure very close to the line of sight, ~1.5°.

The original goal for obtaining the observations presented here was to better elucidate the nature of LH146, originally identified as a galaxy cluster at z = 1.753. We confirmed its redshift and found three more galaxies with very similar values. So there is a galaxy cluster or protocluster at z = 1.75, but LH146 may not be it. We were not able to completely achieve our original goal because the new data have instead revealed that LH146 is in a region containing several components whose relation to each other is not clear. It is likely that we would require a similar set of data for each component to make further progress, which will be a daunting task given their redshifts.

We thank Fabio Bresolin, Andy Fabian, Brian McNamara and Tian-Tian Yuan for discussions that helped improve this paper. JPH thanks Herschel grant 1438631 from the Jet Propulsion Laboratory for support.

*Facilities*: *Subaru* (MOIRCS, FMOS), *Spitzer* (IRAC), XMM-Newton

**Table 1**
Journal of SUBARU Spectroscopic Observations

| Date (UT) | Spectrograph | Exposure (900 s) | Seeing (″) |
|---|---|---|---|
| 13 Mar 2011 | MOIRCS | 18 | 0.5 |
| 14 Mar 2011 | MOIRCS | 12 | 0.4 |
| 30 Dec 2012 | FMOS H-short | 14 | 0.9 |
| 31 Dec 2012 | FMOS H-long | 14 | 1.3 |

**Table 2**
Properties of Spectroscopically Confirmed Galaxies

| ID | α(2000) | δ(2000) | $z_p$ | $z_s$ | J | [3.6] | Features |
|---|---|---|---|---|---|---|---|
| 206708 | 163.3524 | 57.3970 | 1.769 | 1.753 | 22.02±0.09 | 20.03±0.08 | Ab: K, H, Hβ, NaI D |
| 87566 | 163.3437 | 57.3840 | 1.730 | 1.685 | 22.13±0.09 | 20.05±0.07 | Ab: 4000Å break |
| 88512 | 163.3551 | 57.3900 | 1.730 | 1.695 | 23.49±0.31 | 22.47±0.22 | Em: [OII], Hβ [OIII], Hα |
| 87433 | 163.3570 | 57.3830 | 1.752 | 1.691 | 21.88±0.08 | 20.58±0.08 | Em: [OII], Hα, [NII] |
| 87177 | 163.3752 | 57.3814 | 1.749 | 1.677 | 21.74±0.07 | 20.11±0.08 | Em: Hβ, Hα, [NII] |
| 89200 | 163.3507 | 57.3938 | 1.731 | 1.675 | 23.24±0.25 | 21.16±0.12 | Em: Hα, [NII] |
| 94310 | 163.2995 | 57.4241 | 1.698 | 1.755 | 22.15±0.10 | 20.79±0.13 | Em: [OII], Hβ |
| 88683 | 163.3111 | 57.3910 | 2.064 | 1.748 | - | 22.31±0.29 | Em: [OII], Hβ, [OIII] |
| 94757 | 163.3492 | 57.4256 | 1.689 | 1.747 | 23.11±0.22 | - | Em: [OII] |

Note: All but ID 89200 have spectra obtained with MOIRCS.

**Table 3**
Properties of Diffuse X-ray Sources If Thermal

| ID | counts[a] | $r_{ap}$[a] $r_{500}$[a] | $f_{500}(0.5,2.0)$[a] $10^{-16}$ erg cm$^{-2}$ s$^{-1}$ | $L_{500}(0.1,2.4)$[a] $10^{43}$ erg s$^{-1}$ | kT keV | $M_{200}$[a] $10^{13}$ M$_\odot$ | $r_{200}$[a] Mpc | σ km s$^{-1}$ |
|---|---|---|---|---|---|---|---|---|
| LH146[b] | 172±36 | 0.80 | 5.53±1.17 | 3.3±0.7 | 1.6±0.1 | 4.5±0.6 | 0.38 | 401 |
| LH146NWX[c] | 189±56 | 1.11 | 4.36±1.32 | 2.8±0.9 | 1.5±0.2 | 4.0±0.7 | 0.39 | 388 |
| LH146SEX[d] | 225±44 | 0.71 | 30.9±8.1 | 11.8±3.1 | 2.8±0.3 | 10.0±1.6 | 0.51 | 526 |

[a] net counts in elliptical regions, $r_{ap}$ is the characteristic size of the region equal to the square root of its semi major times semi minor axes. The flux and luminosity, in the 0.5 - 2.0 keV and 0.1 - 2.4 keV bands respectively, are within $r_{500}$ and the mass is within $r_{200}$, the radii within which the average cluster density is 500 (200) times the critical density at the redshift of the object.
[b] Ellipse 46 in Figure 4, z = 1.75
[c] Ellipse 37 in Figure 4, z = 1.75
[d] Ellipse 48 in Figure 4, z = 1.75

**Table 4**
Properties of Diffuse X-ray Sources If Inverse Compton Scattering of CMB

| ID | counts[a] | $r_{ap}$[a] arcmin | $f(0.5,2.0)$[a] $10^{-16}$ erg cm$^{-2}$ s$^{-1}$ | $L(2,10)$[a] $10^{42}$ erg s$^{-1}$ | Γ | Length[a] kpc |
|---|---|---|---|---|---|---|
| LH146NWX[b] | 189±56 | 0.49 | 5.50±1.63 | 13.6±4.03 | 2.0 | 785 |
| LH146SEX[c] | 225±44 | 0.41 | 6.53±1.28 | 1.11±0.22 | 2.0 | 520 |

[a] net counts in elliptical regions, $r_{ap}$ is the characteristic size of the region equal to the square root of its semi major times semi minor axes. The flux and luminosity are in the 0.5 - 2.0 keV and 2.0 - 10 keV bands respectively. Length is the physical size of the longest dimension.
[b] Ellipse 37 in Figure 4, z = 1.75 assumed
[c] Ellipse 48 in Figure 4, z = 0.61

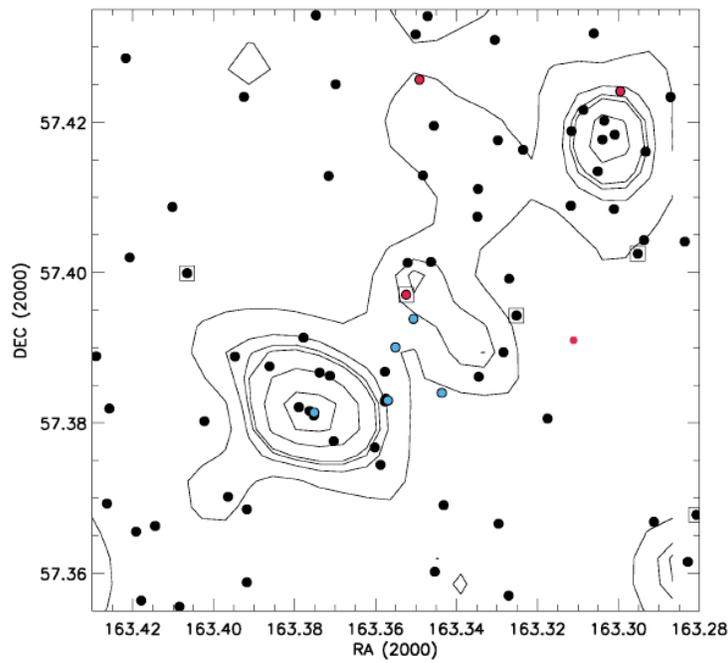

**Figure 1.** Left. Dots are galaxies with photometric redshifts between 1.65 and 1.80 and with errors ≤ 0.1. Blue and red dots are galaxies with spectroscopic redshifts of 1.675 - 1.700 and 1.745 - 1.755, respectively. Squares mark quiescent galaxies. Isopleths are at 4.2, 8.6, 10.5, 11.5, 15.3 and 19.2 galaxies arcmin$^{-2}$. The background is 2.1 galaxies arcmin$^{-2}$.

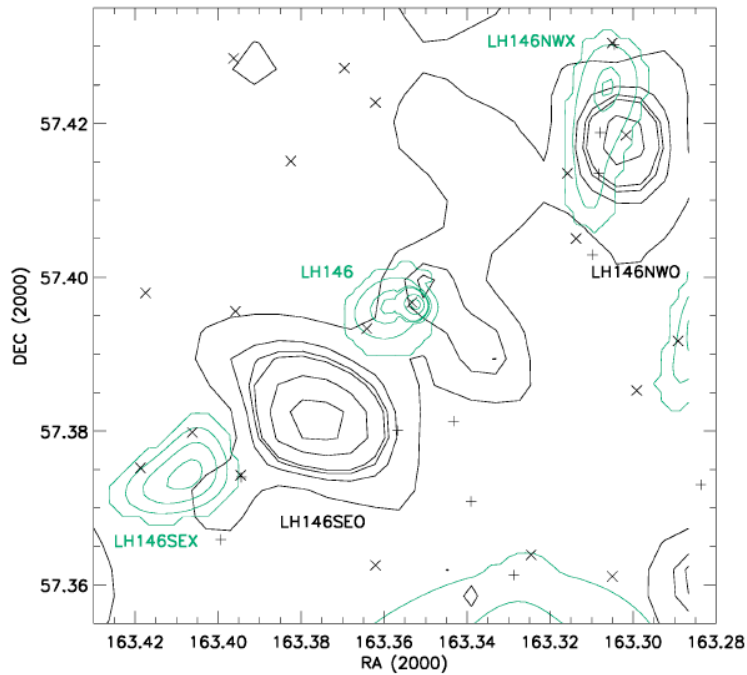

**Figure 1.** Right. The isolpleths are the same as in Figure 1 left. Exes are X-ray sources from Brunner et al. (2008), which are mostly point sources. Crosses are radio sources from Biggs & Ivison (2006). Green contours are the wavelet reconstruction of diffuse X-ray sources. Their contours, in units of $10^{-16}$ erg cm$^{-2}$ s$^{-1}$ arcmin$^{-2}$ above background in the 0.5-2.0 keV band, are: 1.2, 3.0, 3.5, 3.7, 4.4, and 4.9 (LH146); 1.2, 2.4, 3.0, and 3.1 (LH146NWX); 1.2, 3.0, 3.6, 4.1 (LH146SEX); 1.8, 2.6, and 3.0 (West source); 8.4 and 24 (South source).

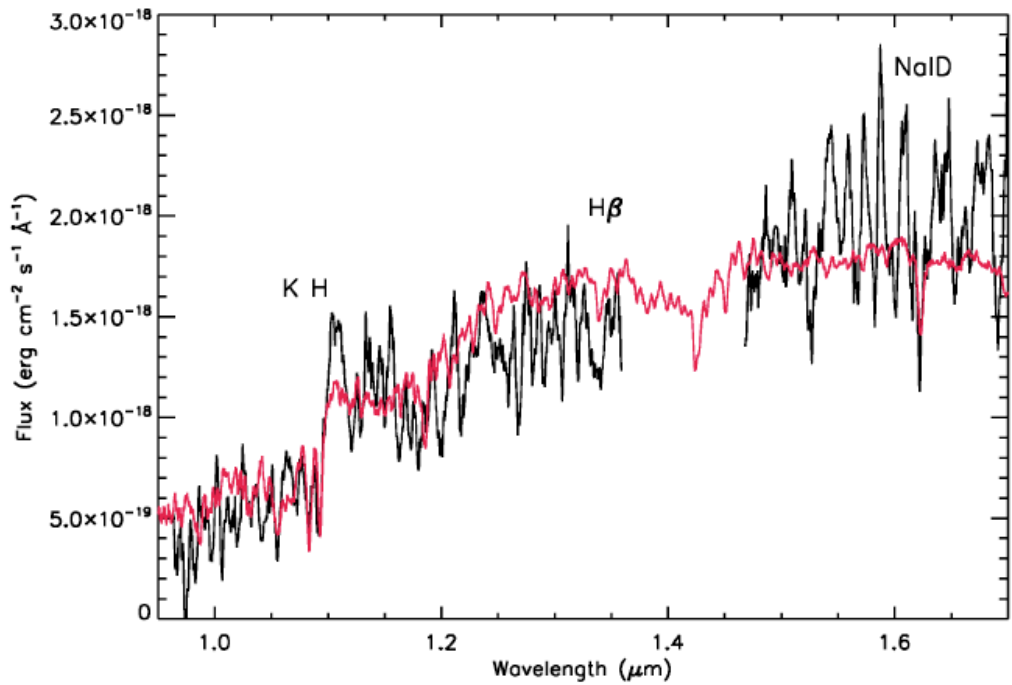

**Figure 2.** Smoothed spectra of objects in Table 2. Absolute fluxes come from scaling to the UKIDSS J magnitudes. The Hα and [NII] lines are resolved in the MOIRCS data, but not at the resolution of the plots. We do not show the region of reduced atmospheric transparency between the J and H bands due to its high noise. One of the Hβ emission and the two [OIII] λ5007 lines noted in Table 2 are at the edge of or in that region and, although visible in the 2 dimensional spectral data, are not apparent in the one dimensional plots here. a. ID 206708 where we have summed the data presented here with that in H10 for a total exposure of 52,800 s. The red line is the average elliptical galaxy spectrum from Eisenstein, Hogg and Fukugita (2003).

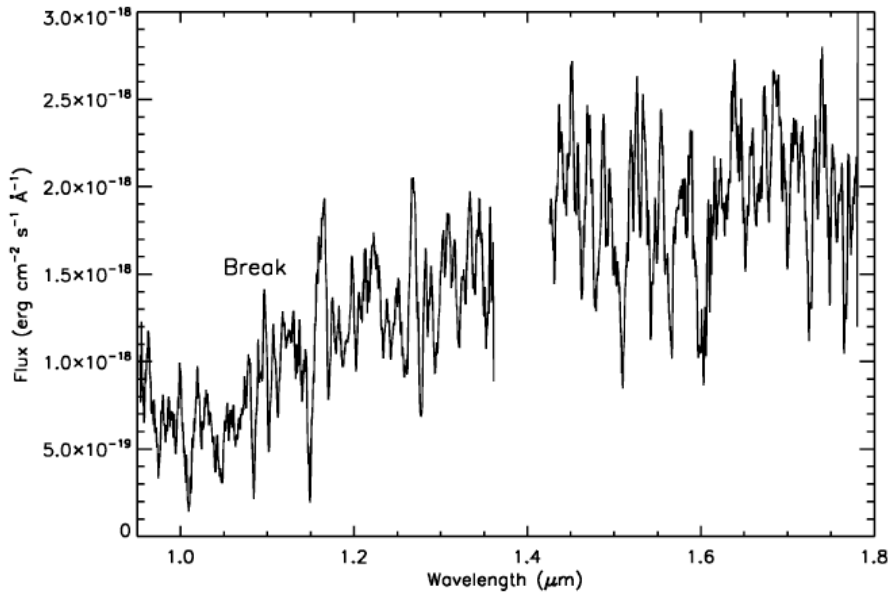

**Figure 2b.** ID 87566

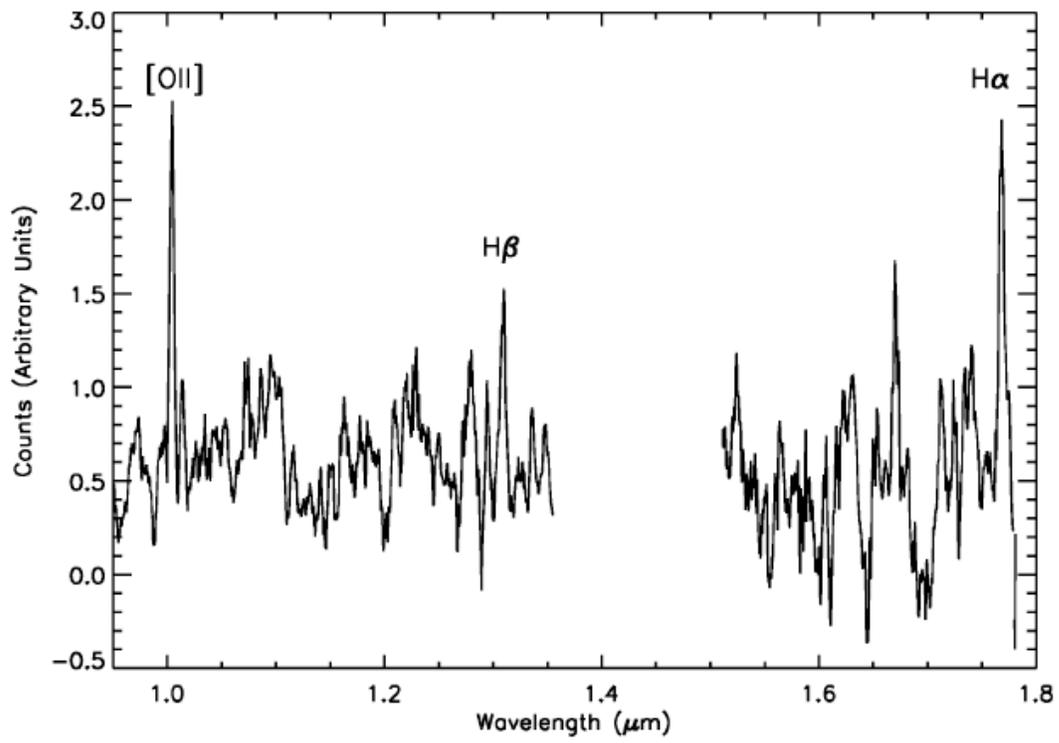

**Figure 2c.** ID 88512

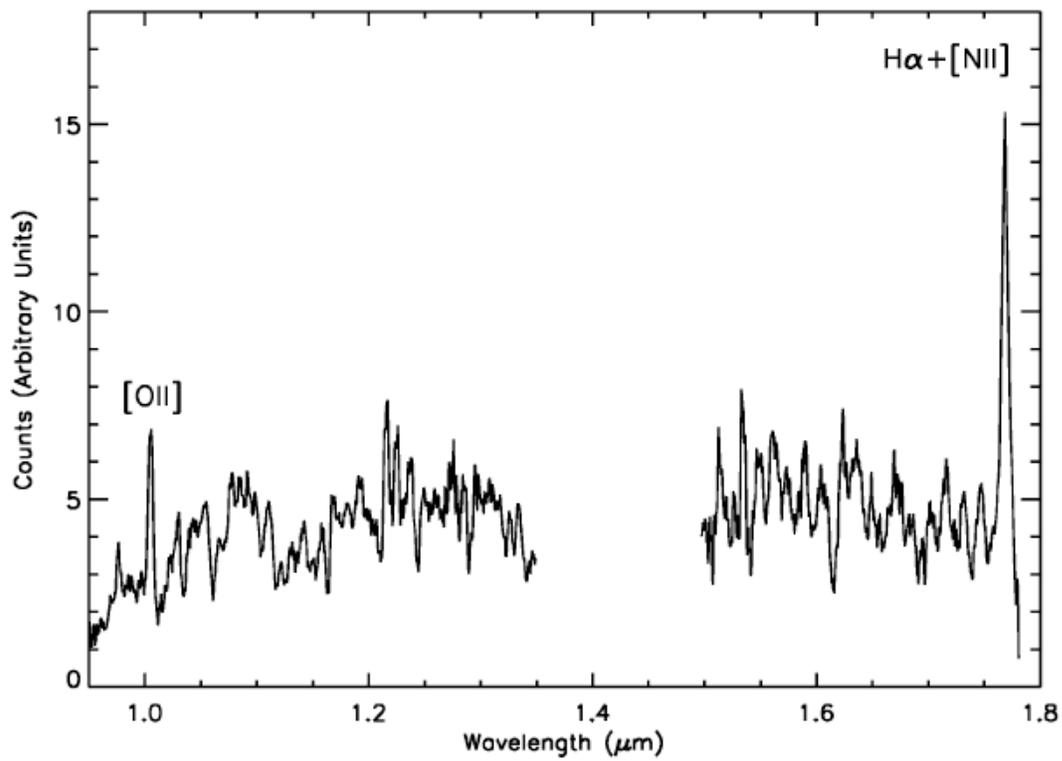

**Figure 2d.** ID 87433, the [OII] λ3727 and [NII] extend ±0.9″ above and below the continuum with no detectable velocity spread. Hα is partially obscured by telluric absorption.

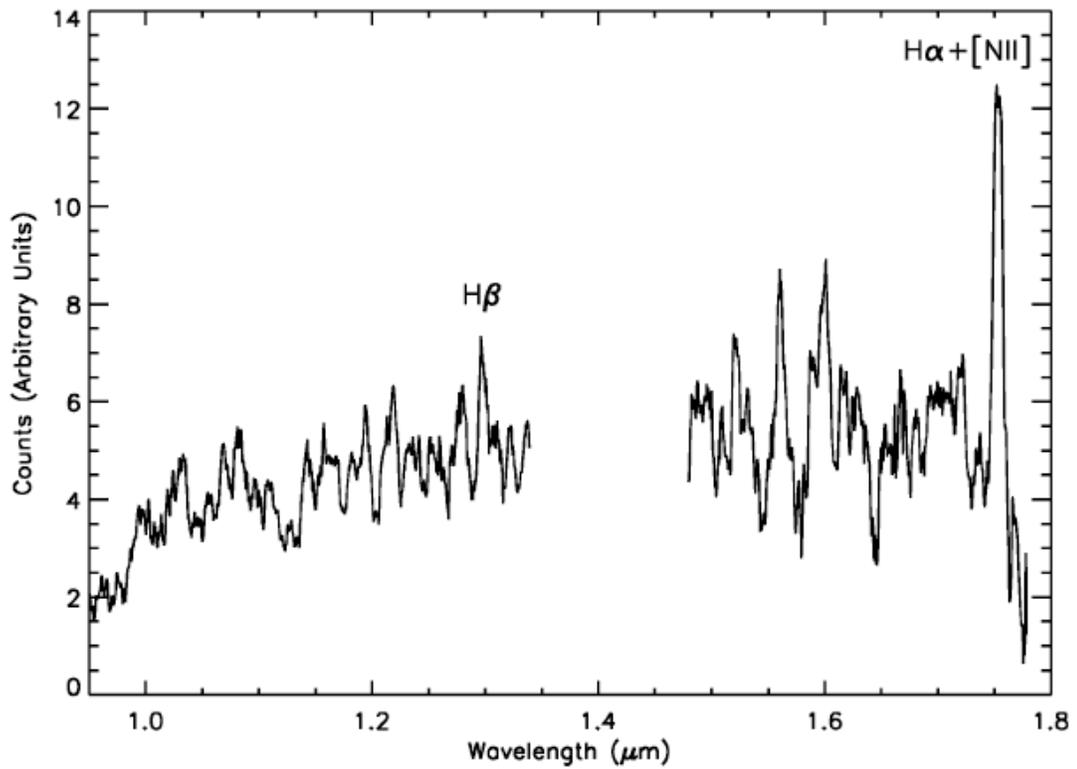

**Figure 2e.** ID 87177, the Hα and [NII] extend ±1.3″ above and below the continuum and their velocity spreads are ±191 km s$^{-1}$.

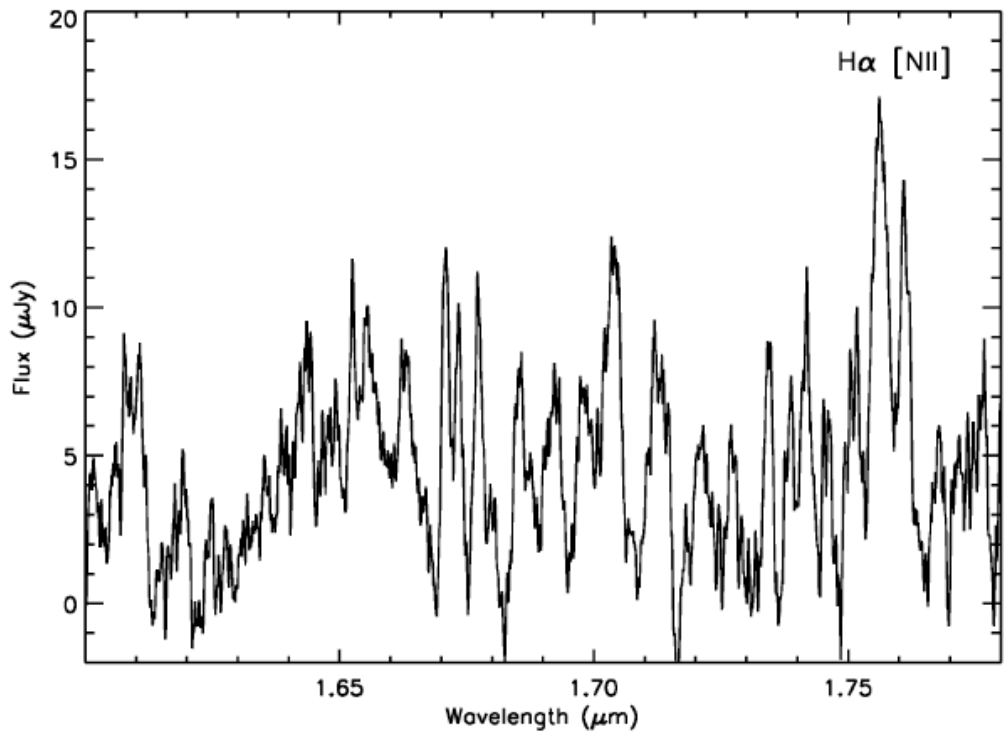

**Figure 2f.** ID 89200, FMOS data.

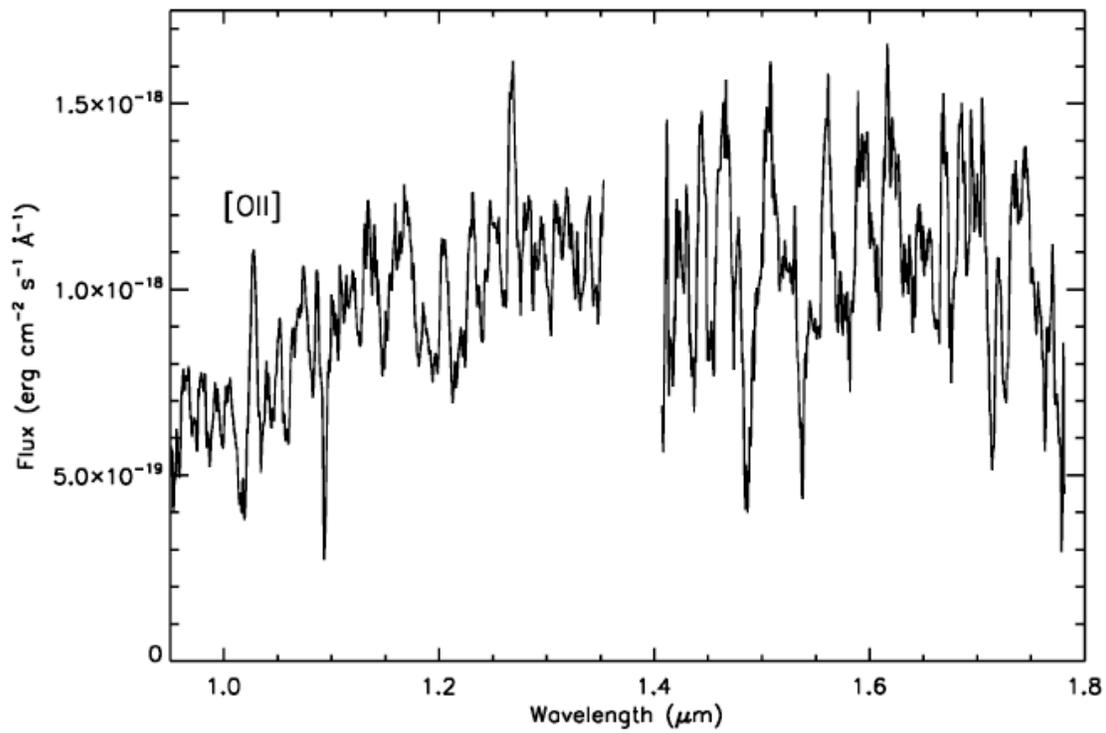

**Figure 2g.** ID 94310

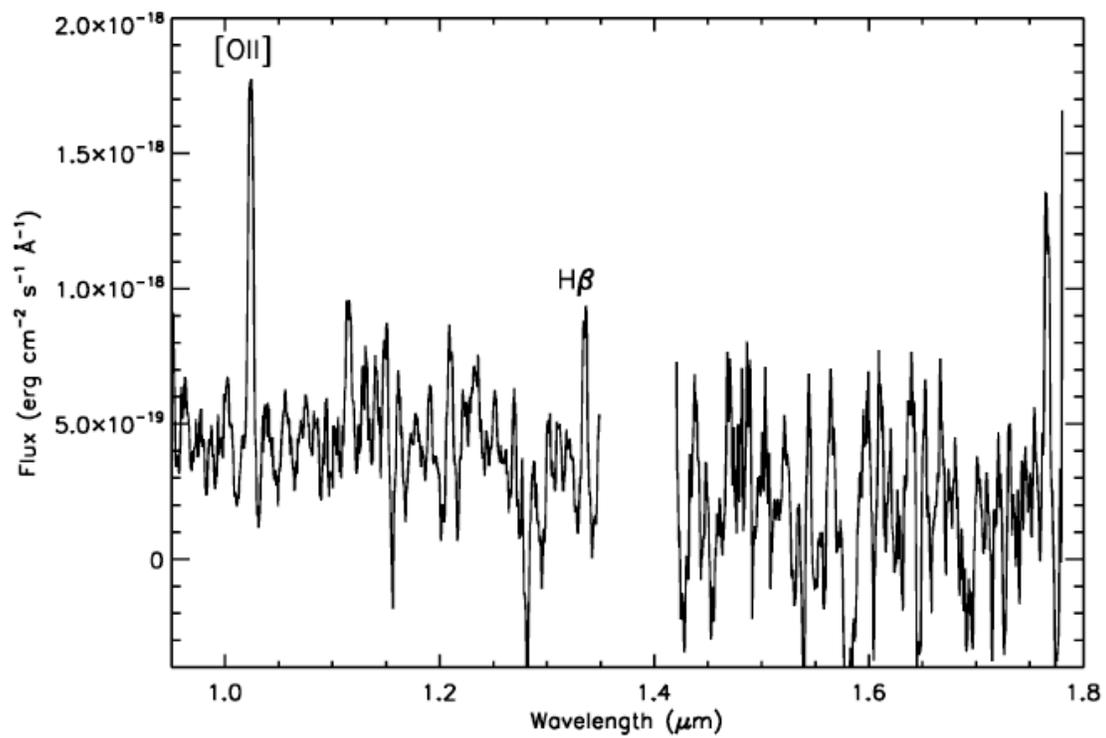

**Figure 2h**. ID 88683 this object is below the UKIDSS J magnitude limit so we used the average scaling from instrumental to absolute fluxes, which is uncertain by at least a factor of 2.

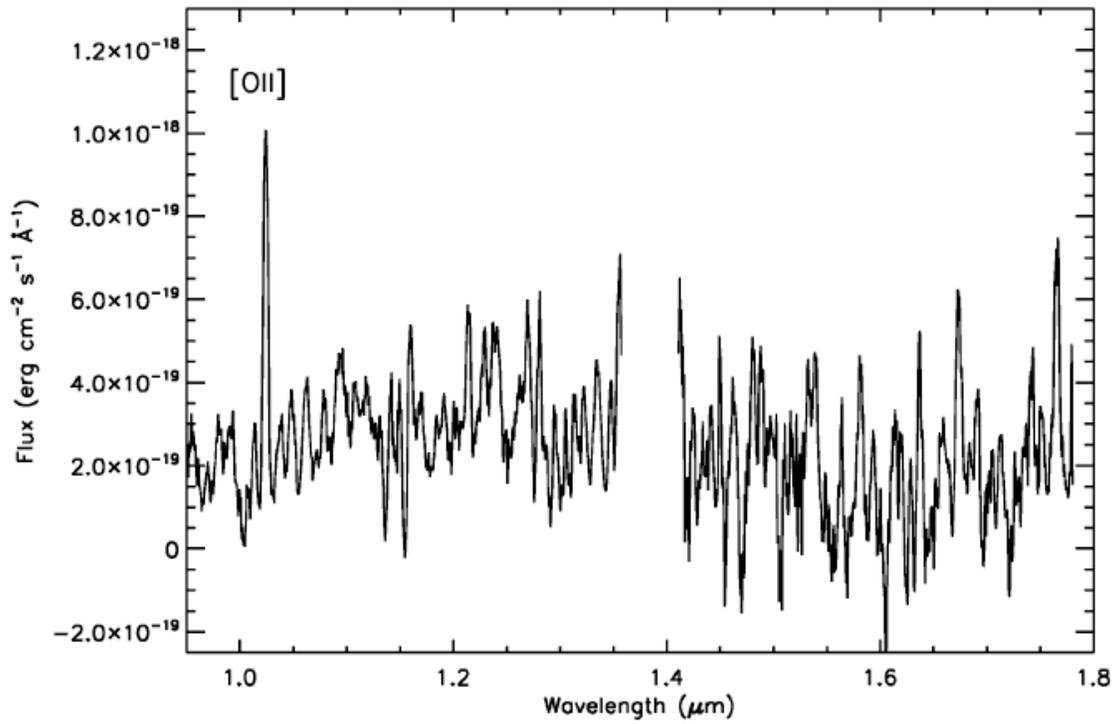

**Figure 2i.** ID 94757

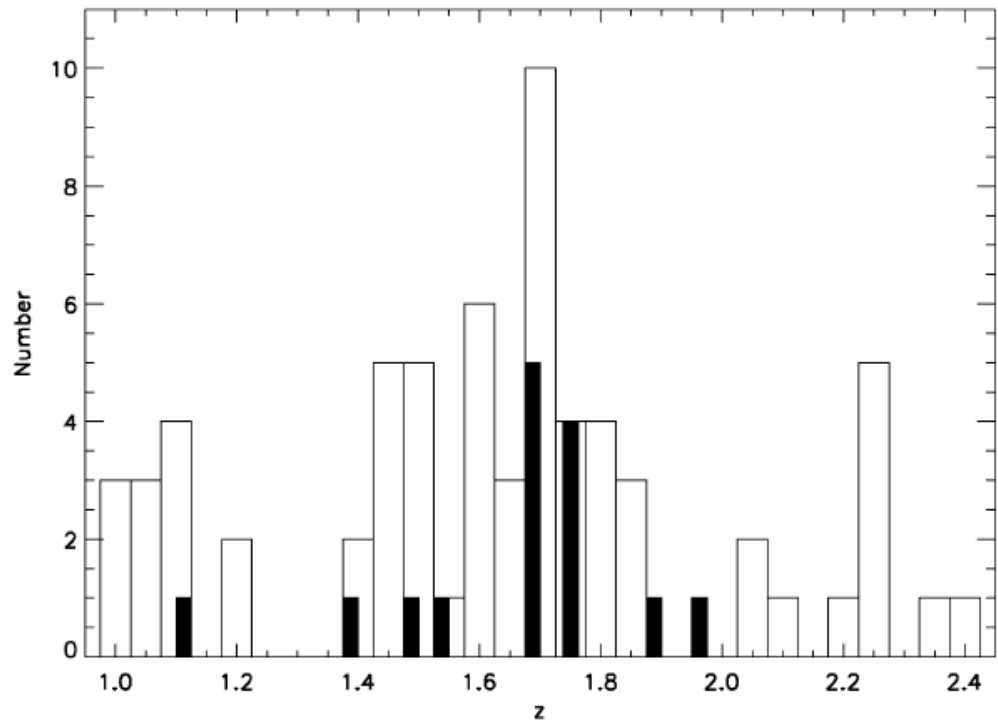

**Figure 3.** Redshift histograms. The open histogram shows photometric redshifts with errors ≤ 0.1 and within square boxes with 1′ sides centered on LH146SEO and LH146NWO. The filled histogram shows spectroscopic redshifts in the field of Figure 1 from the observations described in section 3.1. The bins are 0.05 for photometric redshifts and 0.025 for spectroscopic redshifts.

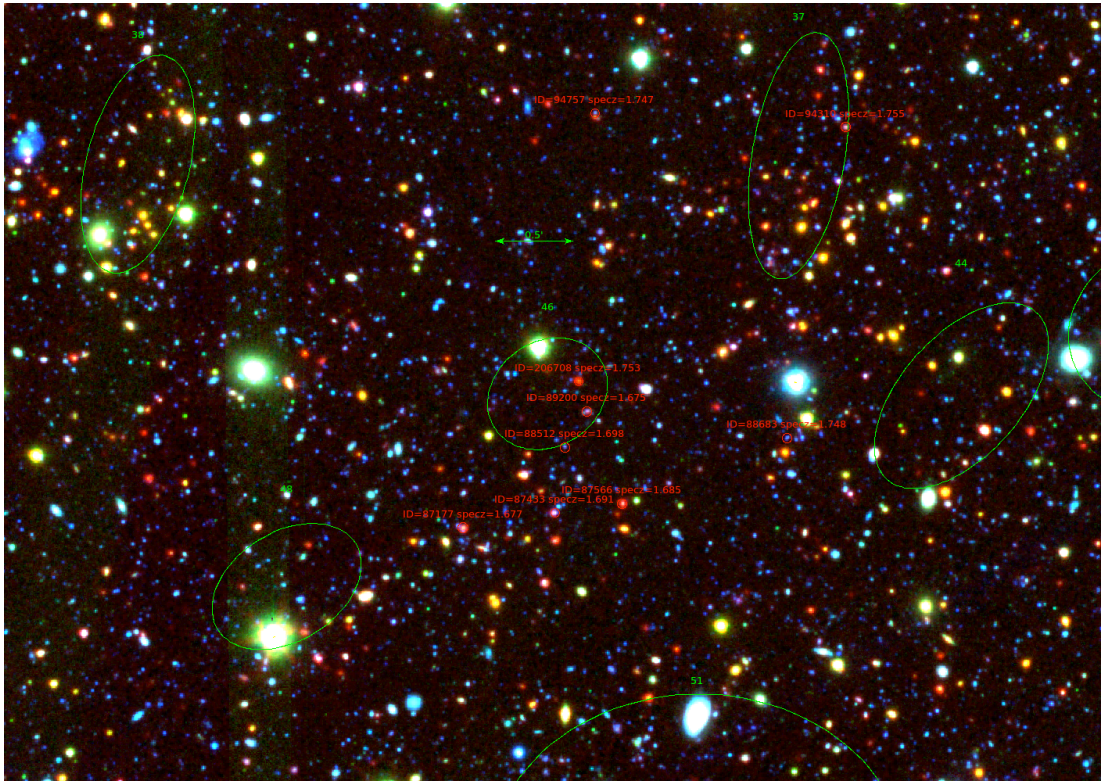

**Figure 4.** RGB image centered on LH146 composed of a 19,972 s $B_J$ exposure from the LBT, a 10,400 s z′ exposure from the Subaru telescope and a 620 s [3.6] exposure from Spitzer. The green ellipses numbered 37, 46 and 48 are the extraction regions for the counts from the diffuse X-ray sources in Tables 3 and 4. Other ellipses denote other diffuse X-ray sources in the field. Confirmed large scale structure galaxies are circled and their IDs and spectroscopic redshifts are given. The scale bar near the center is 0.5′ long.

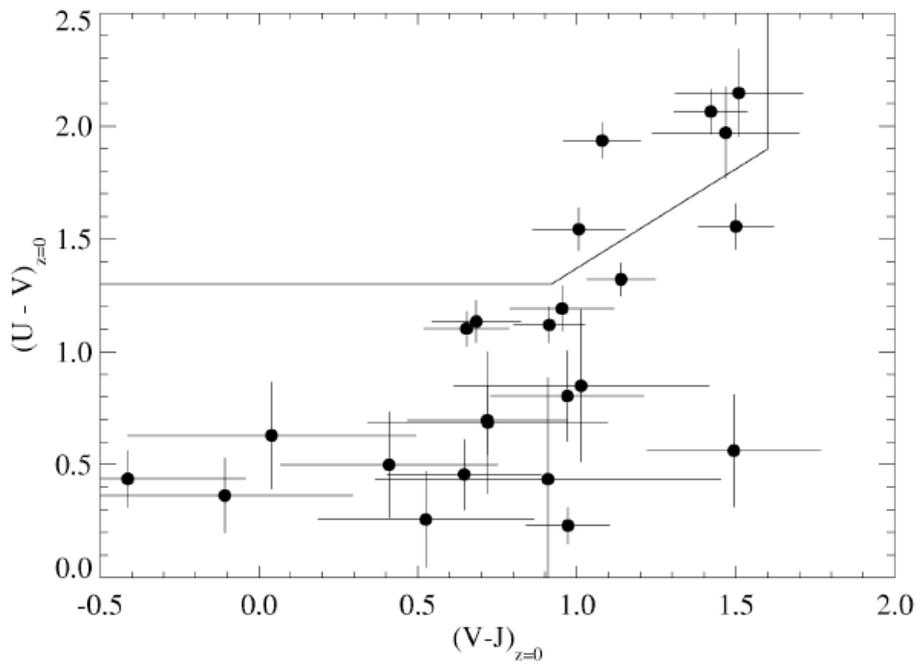

**Figure 5.** Interpolated rest frame color-color diagram for all galaxies in Figure 1 left detected in the three needed bands. The quiescent elliptical galaxies are the five objects above the line.

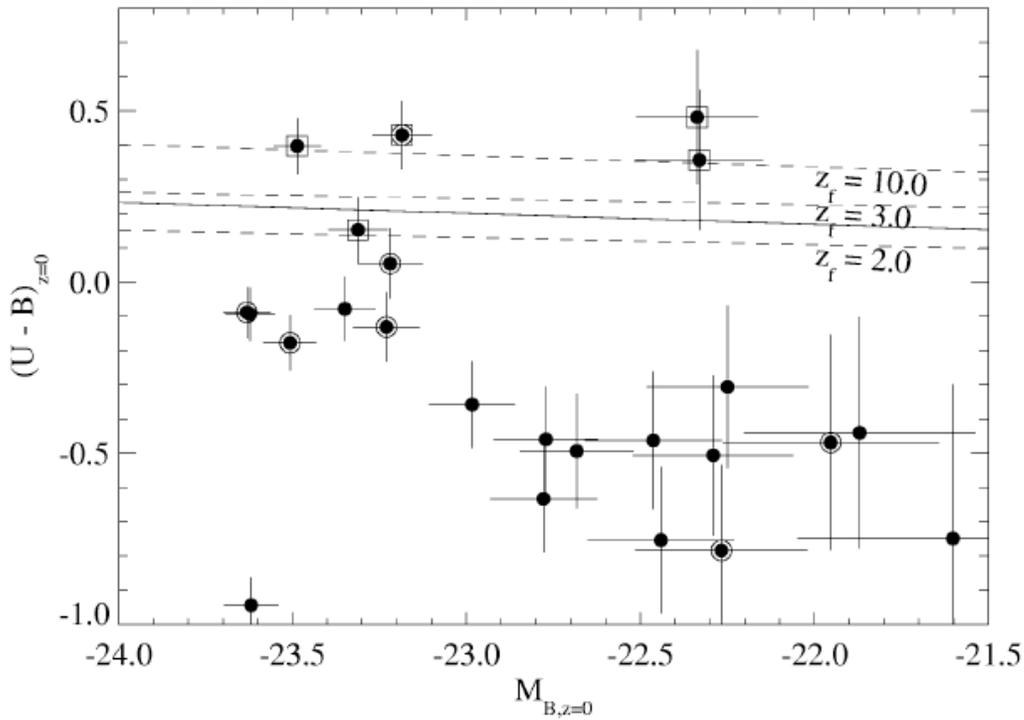

**Figure 6.** Rest frame color-magnitude diagram in the Vega system. Points enclosed by boxes are the five galaxies identified as quiescent ellipticals in Figure 5. Points circled are spectroscopically confirmed members. The solid line is the red sequence in the same bands of ClG J0218.3-0510 at $z = 1.62$ from Papovich et al. (2010). The dashed lines are theoretical red sequences of galaxies that formed at $z = 2, 3,$ and $10$.